\documentclass[12pt,preprint]{aastex}

\shorttitle{STRUCTURAL ANALYSIS OF EARLY-TYPE GALAXIES}
\shortauthors{A.W.\ Graham, et al.}

\begin{document}

\title{A New Empirical Model for the Structural Analysis of
Early-type Galaxies and a Critical Review of the Nuker Model\altaffilmark{1}}
\author{Alister W.\ Graham}
\affil{Department of Astronomy, University of Florida, Gainesville, FL, USA}
\email{Graham@astro.ufl.edu}

\author{Peter Erwin, I.\ Trujillo, and A.\ Asensio Ramos}
\affil{Instituto de Astrof\'{\i}sica de Canarias, La Laguna, E-38200, Tenerife, Spain}

\altaffiltext{1}{Based on observations made with the NASA/ESA {\sl
Hubble Space Telescope}, obtained at the Space Telescope Science
Institute, which is operated by the Association of Universities for
Research in Astronomy, Inc., under NASA contract NAS 5-26555.}

\begin{abstract}
The {\it Nuker law} was designed to match the inner few 
($\sim$3-10) arcseconds of predominantly nearby ($\lesssim$ 30 Mpc) 
early-type galaxy light-profiles; it was never intended to describe 
an entire profile.  The S\'ersic model, on the other hand, was developed 
to fit the entire profile; however, due to the presence of partially 
depleted galaxy cores 
the S\'ersic model cannot always describe the very inner region.   
We have therefore developed a new empirical model consisting of an inner 
power-law, a transition region, and an outer S\'ersic model 
to connect the inner and outer structure of elliptical galaxies.  
We have additionally explored the stability of the Nuker model parameters. 
Surprisingly, none are found to be stable quantities; 
all are shown to vary 
systematically with a profile's fitted radial extent, and often 
by more than 100\%. 
Considering elliptical galaxies spanning a range of 7.5 magnitudes, 
we reveal that the central stellar densities of the underlying host 
galaxies {\it increase} with galaxy luminosity 
until the onset of core formation, detected only in the brightest 
elliptical galaxies.  
We suggest that the so-called ``power-law'' galaxies 
may actually be described by the S\'ersic model over their entire 
radial range.  
\end{abstract}


\section{Introduction}

Early {\it Hubble Space Telescope} ({\it HST}) observations of elliptical galaxies 
and the bulges of disk galaxies, hereafter collectively referred to as `bulges' 
(Crane et al.\ 1993; Ferrarese et al.\ 1994; Jaffe et al.\ 1994; 
%
%
Forbes, Franx, \& Illingworth 1995), confirmed the 
ground-based conclusions of Kormendy (1985) and Lauer (1985):
Galaxy models with flat cores (e.g., King's 1966 model) 
do not describe the majority of elliptical 
galaxies, or at least the resolved part of the profile.  
In almost all galaxies surveyed, the surface brightness 
profile continued to rise inward until resolution was lost. 
Subsequently, with the enhanced image quality afforded by the 
{\it HST} have come new models to describe the centers of nearby bulges. 

Ferrarese et al.\ (1994) introduced two classes of galaxies according 
to the behavior of the inner surface brightness profile.  
Those with a resolved core flattening towards 
the center were labeled as ``Type I'', and those that roughly follow 
a steep power-law all the way into the center were designated as 
``Type II'' galaxies.  These authors introduced a 4-parameter 
double power-law model to quantify the shape of the galaxy 
profile within the inner $\sim 10\arcsec$.  A core radius marked the 
transition between the inner 
and outer power-laws having slopes $\beta_1$ and $\beta_2$, respectively.  
All of their ``Type I'' galaxies had an inner slope shallower than -0.31 
(none of them had a slope of zero); all but one of their ``Type II'' 
galaxies had a slope steeper than -0.47.  

Modeling a larger galaxy sample, Lauer et al.\ (1995; 
see also Kormendy et al.\ 1994) 
confirmed the above result\footnote{It should be noted that
Rest et al.\ (2001) and Ravindranath et al.\ (2001) have now found several 
high-luminosity galaxies with inner profile slope $0.3<\gamma<0.5$.} 
 --- though they interpreted it differently --- and 
introduced a model with an additional parameter ($\alpha$), which 
better controlled the transition between the two power-laws.  
This model\footnote{This 
5 parameter double power-law model was independently introduced 
by Zhao (1996) to model the spatial (i.e.\ not projected) density 
profiles of elliptical galaxies (see also Zhao 1997).} 
was designated the `Nuker law' by these authors.  It can be 
written as 
\begin{equation}
I(r)=I_b2^{(\beta-\gamma)/\alpha}\left(\frac{r}{r_b}\right)^{-\gamma}
\left[1+\left( \frac{r}{r_b} \right)^{\alpha}\right]^{(\gamma-\beta)/\alpha}.
\end{equation}
The intensity at the core radius, also known as the break-radius 
$r_b$, is denoted by $I_b$.  The inner power-law slope is now denoted 
by $\gamma$ and the outer power-law slope is denoted by $\beta$. 
This model reduces to the form proposed by Ferrarese et al.\ (1994)
when $\alpha=2(\beta_2-\beta_1)$.  
Lauer et al.\ (1995) refer to 
galaxies with $\gamma<0.3$ as ``core'' galaxies and galaxies with 
$\gamma>0.5$ as ``power-law'' galaxies.  

Application of the Nuker model has proved extremely popular, and 
there are physical grounds to interpret the reduction in central profile 
slope and the implied core depletion. 
Many authors have discussed how the inner region of a galaxy 
may have been partially evacuated by the coalescence of merging 
supermassive black holes (SMBHs; e.g., Ebisuzaki, Makino, \& Okumura 1991; 
Makino \& Ebisuzaki 1996; Faber et al.\ 1997; Makino 1997; 
Quillen, Bower, \& Stritzinger 2000; Alexander \& Livio 2001; 
Milosavljevic \& Merritt 2001). 
Conversely, the presence of ``power-law'' cusps has been used
to argue for adiabatic growth of central black holes, with the growing
black hole reshaping the central region (e.g., van der Marel 1999), although
Ravindranath, Ho, \& Filippenko (2002) have used the fitted power-law slopes
to argue against this scenario. 
To better understand galaxy ``cores'', one would like to measure changes 
to a galaxy's inner profile relative to its original shape.  

Recently, the overall  
``shape'' of a bulge's light-profile (as parameterized by the S\'ersic 
1968 $r^{1/n}$ shape index $n$) has been shown to 
correlate strongly ($r_s=0.92$) with the mass of its central 
SMBH (Graham et al.\ 2001a, 2002; Erwin, Graham, \& Caon 2003).  
This implies a strong connection between the formation and structure 
of the entire bulge and the formation of the central black hole.   
The central regions of bulges are thus directly related to 
the global bulge structure, and so one would like 
to connect these two regimes.  

The Nuker model -- with 5 free parameters -- can only describe the 
inner light-profile of a bulge; it was never designed to model an entire 
profile and is thus unable to make a connection between the inner 
profile and the overall bulge structure\footnote{A possible exception
to this remark comes from the observation that the outer light-profiles
of the more massive brightest cluster galaxies can be well approximated
by a power-law (e.g.\ Graham et al.\ 1996).  It should perhaps further
be noted that the S\'ersic model tends towards a power-law for 
large values of $n$.}.  
The S\'ersic model -- with 3 free parameters -- matches 
the entire radial extent of most bulge light-profiles 
remarkably well, with the exception of the inner few arcseconds 
for some galaxies. 
By joining, at the break radius, an outer S\'ersic profile 
with an inner power-law, one might hope to be able to describe 
the complete light-profiles of bulges when the S\'ersic model 
alone is inadequate. 

This issue will be addressed here and in a companion paper 
(Trujillo et al.\ 2003, Paper II).  To do this, 
the merits of the individual Nuker parameters will first be explored 
in Section 2. 
Section 3 then describes the S\'ersic model, which, for a number of 
illustrative purposes, is applied here to the central, early-type 
galaxy light-profiles presented in Lauer et al.\ (1995).  
Given the short-comings of both models to describe the 
complete light-profiles of all bulges, a new empirical model is 
introduced in Section 4 
and is illustrated with application to both a ``power-law'' and a 
``core'' galaxy profile.   In Paper II, we 
apply the new model to radially complete profiles from a larger sample of 
early-type galaxies.  
Correlations between the global and core properties will be presented 
in a forthcoming paper.  A recapitulation of the main points in this paper 
are provided in Section 5.

\section{The Nuker model}

In those bright galaxies where Faber et al.\ (1997; and references 
therein) detected a ``core'', the break radius $r_b$ and the intensity at
this radius ($I_b$) are thought to denote the onset of a physical 
transition in a galaxy's profile.  
Together with the central velocity 
dispersion, Faber et al.\ (1997; see also Faber et al.\ 1987) constructed 
a `core fundamental plane', from which they concluded that: cores are 
in dynamical equilibrium; $r_b$ and $I_b$ are meaningful dynamical 
parameters [at least in the case of ``core'' galaxies]; velocity anisotropy 
does not vary greatly among ``core'' galaxies; for most galaxies, the
mass of any central SMBH does not dominate the core potential; and 
the core $M/L$ ratio varies smoothly over the fundamental plane. 

Faber et al.\ (1997) also noted, however, that the value of $r_b$ 
[and hence $I_b$] is not robust for ``power-law'' galaxies.  
The difficulty the Nuker model has in obtaining stable parameters 
for such galaxies, and hence some quantity which reflect some 
fixed physical structure, is a consequence of their smooth continuously 
curving profiles which have no obvious core. 

The S\'ersic model has a smooth, continuous profile that in fact 
resembles the observed ``power-law'' profiles. 
The S\'ersic model has also recently been shown to provide a 
good description to both the outer {\it and inner} profiles of 
{\it HST}-resolved, low-luminosity elliptical galaxies (after 
accounting for the central excess flux; Stiavelli et al.\ 2001; 
Graham \& Guzm\'an 2003). 
Consequently, we explore the natural question: Are the so-called 
``power-law'' galaxies simply the bright end of these S\'ersic--$r^{1/n}$ 
galaxies without (resolved) break radii, depleted cores, or true central 
power-laws?  
To further this idea we will look at a 
compilation of galaxies spanning a large range in absolute magnitude. 

Figure~\ref{fig1} shows the central surface brightnesses of the Nuker 
team's elliptical galaxies plotted against their absolute magnitudes. 
The absolute $V$-band magnitudes were obtained from Faber et al.\ 
(1997; their Table 2) and converted back into $B$-band magnitudes 
using their ($B-V$) color term (their Table 1). 
The central surface brightnesses shown here are those of the 
Nuker model 
at $r=0.1\arcsec$, corrected for Galactic extinction, 
and converted to the $B$-band in the 
same manner as done for the magnitudes\footnote{When no ($B-V$) color term was
given by Faber et al.\ (1997), a value of 0.9 has been adopted.  
This may lead 
to slightly increased scatter in Figure~\ref{fig1}.}. 
Faber et al.\ (1997) wrote that ``severe nuclei were ignored in 
fitting Nuker laws''.  This can be seen in the Nuker models fitted to both the 
``core'' (e.g., NGC~6166 and A2052) 
and ``power-law'' galaxies in Byun et al.\ (1996).  
It is noted that 
the overall placement of the Nuker team's galaxies in Figure~\ref{fig1} is 
the same when the actual surface brightness within the inner 0.1$\arcsec$, as 
given in Table 4 of Lauer et al.\ (1995), is used.  
Surface brightness values from Nuker models extrapolated to $r=0$ 
were not used because of this model's power-law behavior, 
and hence overestimation of the true (finite) central surface brightness. 

Added to this diagram are the Virgo and Fornax
elliptical galaxies imaged by Caon, Capaccioli, \& D'Onofrio (1993) and 
D'Onofrio, Capaccioli, \& Caon (1994).  The S0 galaxies 
have again been excluded as we are interested in the properties 
of bulges (not bulges and disks combined).  
The observed, model-independent central surface 
brightness\footnote{It should be noted that the central surface 
brightnesses of these galaxies were obtained with ground-based 
CCD imaging under $\sim1\arcsec$ seeing; the true central surface 
brightness is therefore brighter than shown here.} 
and model-independent magnitude of these elliptical galaxies  
are shown; there has been 
no recourse to the S\'ersic model to determine these values, 
although readers are reminded 
that these galaxies were very well fitted with a S\'ersic model 
by the above authors (see also D'Onofrio 2001).  
The dwarf elliptical galaxies in Binggeli \& Jerjen (1998; their Table 1) 
are also included in Figure~\ref{fig1}.  These authors also avoided 
the nuclear point sources 
%
%
and so their central surface brightness measurements are from a 
S\'ersic model fitted to the 
underlying bulge (extrapolated to $r = 0\arcsec$), 
uncontaminated by possible nuclear star clusters.  
%
%
Finally, the dwarf elliptical galaxies imaged with the {\it HST} by 
Stiavelli et al.\ (2001) are shown.  
The magnitudes are from their table 1, and the central surface brightness
values have been read off from the S\'ersic fits in their figure 1 (these
readings are probably accurate to 0.05 mag arcsec$^{-2}$).  
The S\'ersic fits were also made after excluding any excess central flux.  
Both the magnitude 
and central surface brightness measurements from Stiavelli et al.'s galaxy 
sample have been converted here from the $V$-band to 
the $B$-band assuming a constant ($B-V$) color of 0.9.  

 
It should also be understood that the {\it total}, central surface 
brightnesses of many galaxies included in Figure~\ref{fig1} 
are brighter than what is shown there, what is shown are 
estimates of the central surface brightness values of the {\it underlying 
host} galaxies.   With this understanding, the central galaxy intensity is 
seen to increase with bulge luminosity (see also Caldwell 1983)  
such that $M_B \propto -(3/2) \log(I_{0,B})=(3/5)\mu_{0,B}$ 
(Figure~\ref{fig1}), 
but above a certain threshold ($M_B \lesssim -20.5$) one observes a 
reversal of this trend.  The bright end of this trend can be seen in 
Phillips et al.\ (1996, their figure 6) and Faber et al.\ (1997, their 
figure 4c).   {\it Hot galaxies 
have projected central stellar 
densities which increase with galaxy luminosity/mass until core 
formation occurs and a break in the light-profile is detected.} 
Our finding would therefore appear to disagree with the interpretation 
by Faber et al.\ (1997), who wrote, ``A major conclusion is that small 
hot galaxies are much denser than large ones''.  They attributed the 
observed reduction in central stellar intensity as one progressed to 
magnitudes fainter than $M_B\sim -20.5$ as a resolution effect, using 
the rare `compact elliptical' galaxy M32 to support this view.  

Past modeling of M32's ground-based light-profile excluded the 
excess flux observed over the inner 10--15$\arcsec$ (e.g., Kent 
1987; Choi, Guhathakurta, \& Johnston 2002).  
If M32 was at the distance of the Virgo cluster,  
this central excess would show up only within the inner 1$\arcsec$ 
and the galaxy would likely be considered `nucleated'.  Excluding 
this unusually sharp core in M32 (Schweizer 1979 and 
Tonry 1984), Graham (2002a) found the 
underlying bulge component has an exponential-like profile; that is, it 
has a relatively shallow inner slope, and he 
derived a central surface brightness 
of 15.31 $R$-mag arcsec$^{-2}$ for the underlying bulge.  
%
%
This roughly translates into a central $V$-band surface brightness of
$\sim$15.7 mag arcsec$^{-2}$ and places it in better agreement with the 
other ``power-law'' galaxies in Figure 4c of Faber et al.\ (1997).  
In any case, because there is evidence suggesting M32 contains an 
outer envelope/disk (Graham 2002a), it has not been included in our 
Figure~\ref{fig1}.

It is noted that the ``power-law'' galaxies in Figure~\ref{fig1} (considered 
to be intermediate-luminosity elliptical galaxies, Faber et al.\ 1997) 
form a continuous extension to the dwarf (low-luminosity) elliptical 
galaxies which are known to be well described by the S\'ersic model 
(Davies et al.\ 1988; Young \& Currie et al.\ 1994; Jerjen, 
Binggeli, \& Freeman 2000).  Excluding the ``core'' galaxies, 
there is no apparent dE/E dichotomy in Figure~\ref{fig1} (see also 
Jerjen \& Binggeli 1997).  
Since the S\'ersic fits to the {\it HST} dwarf elliptical profiles of 
Graham \& Guzm\'an (2003) do an excellent job of describing both the 
outer and the inner profiles (with a point-source used by Graham \& 
Guzm\'an 
to fit those galaxies which are nucleated), and since Jerjen et al.\ 
(2000) were able to fit the highly resolved (due to their proximity) 
Milky Way and M31 dwarf spheroidal profiles at all radii with the 
S\'ersic model, it appears that the ``power-law'' centers of low-luminosity 
ellipticals are simply the inner part of their overall S\'ersic profile.
Moreover, using the S\'ersic-derived (finite) central surface brightness 
values from the Virgo and Fornax galaxies from Caon et al.\ (1993) and 
D'Onofrio et al.\ (1994), galaxies having $M_B \gtrsim -20.5$ overlap 
exactly with the Nuker team's ``power-law'' galaxies (Graham \& Guzm\'an 2003).  
That is, galaxies having 
the same magnitude as the ``power-law'' galaxies have the same central 
surface brightness when derived from the inward extrapolation of the
outer profile's best-fitting S\'ersic model.  

It therefore seems reasonable that the ``power-law'' 
galaxies may indeed simply be S\'ersic--$r^{1/n}$ galaxies without cores. 
This is of interest because it not only helps to 
provide a unifying picture 
of galaxy structure, but reveals that the break radius 
for the so-called ``power-law'' galaxies is not something intrinsically 
physical to these galaxy, 
but is simply a parameter in a model that provides a good reproduction of 
the observed inner light-profile. 
This idea is pursued (and confirmed) by Trujillo et al.\ (2003; see also
Figure~\ref{fig8} in this paper) who model 
the {\it HST} profiles of Lauer et al.\ (1995) and Rest et al.\ (2001)
combined with the outer galaxy profile.   That the S\'ersic
model can describe ``power-law'' galaxies over their entire radial extent, 
is also of interest because 
one replaces 5 parameters which have no clear physical meaning 
with 3 parameters which do, and which fit the entire profile. 
Furthermore, the notion that the inner regions of low-luminosity 
bulges should be treated differently than the outer regions
(i.e., that the inner regions are described by a power-law and the
outer regions by a different function) can be replaced with a 
single unifying model that treats both regions simultaneously.

\subsection{$\alpha$, $\beta$, $\gamma$, and $\gamma\prime$}

The function of the parameter $\alpha$ in the Nuker model (Equation 1) 
is to allow for varying 
degrees of curvature in the surface brightness profile, providing 
a smoother, less abrupt transition between the two power-laws.  
While the transition region is apparently better matched in this way, 
a problem is known to arise when the transition region is apparently 
large and $\alpha$ becomes too small.  When this occurs, 
the slope of the power-law components become 
less and less representative of the observed mean logarithmic slope on 
either side of the break-radius, and more representative of the slope
to the extrapolated model beyond the boundaries of the fitted galaxy profile.  
For smaller values of $\alpha$ (i.e.\ $\lesssim $1), the presence 
of two power-laws often 
fails to emerge; instead, one continuous curving arc 
describes the profile.  As a result, the value of the inner power-law 
slope ($\gamma$) is sometimes zero 
and/or often does not appear to reflect the observed inner slope. 
This aspect of the Nuker model's ability to provide an accurate 
quantification of the observed inner profile slope 
was discussed by Rest et al.\ (2001), 
who also noted the additional difficulty with the Nuker model when the 
break-radius is either smaller than the image resolution, or when 
there simply is no apparent break-radius.  Therefore, in an effort to 
quantify the innermost resolved profile slope, they used an 
additional quantity to accompany the Nuker model.  
Rest et al.\ 
(2001) computed the negative logarithmic slope of the Nuker model at 
0.1$\arcsec$, which they denoted as $\gamma\prime$. 

In practice, one may find that the observed inner profile is well 
approximated by a real power-law and the derivative at 0.1$\arcsec$ matches 
the Nuker model parameter $\gamma$, in which case nothing is gained.
Conversely, one may find that the inner profile is indeed curved 
(or its slope is poorly parameterized by the Nuker model due to a small 
value of $\alpha$) and the local derivative $\gamma\prime$ does not 
match $\gamma$.  One thus derives a new quantity ($\gamma\prime$) 
which is dependent on the radius where it is measured --- that is, when 
the inner profile is not a power-law, $\gamma\prime$ is an apparent, 
rather than absolute, quantity.  If identical galaxies 
are observed at different distances, then they can have different 
values of $\gamma\prime$ ($R=0.1\arcsec$).  This was 
noted by Seigar et al.\ (2002), but not explored or quantified. 

The extent of such changes is illustrated here by computing 
$\gamma\prime(r=0.1\arcsec)$ from the slope of the Nuker model 
(Rest et al.\ 2001, their equation 8) fitted to a range of $r^{1/n}$ 
profiles having $r_e=10\arcsec$ and $n$=1, 2, 3, and 4 (see Section 3).  
By effectively moving each of these galaxy models three times 
further away\footnote{A factor of three in distance corresponds
to the range of distances in the galaxy sample of Carollo \& Stiavelli 
(1998), who computed $<\gamma>$ over 0.1--0.5$\arcsec$.}, 
i.e.\ by simply reducing $r_e$ by a factor of 3, 
$\gamma\prime$ changes from 0.00, 0.09, 0.30, and 0.51 
to 0.13, 0.36, 0.63, and 0.83 for the $n$=1, 2, 3, and 4 models 
respectively.  Thus, initially three galaxies 
would have been classified as core galaxies, but now only one would be
classified as such, even though the actual galaxy structures 
did not change. 
Half of the galaxies in the sample of Rest et al.\ (2001) that are 
classified as core-galaxies using the Nuker model parameter $\gamma$ 
are not classified as core galaxies using the derivative at 0.1$\arcsec$. 
Some of this mismatch is probably due to the distances the galaxies are at, 
and hence what physical radius $\gamma\prime$ was measured at. 

To conclude, $\gamma$ is known to be an extrapolated quantity that does
not reflect the observed inner profile slope of galaxies 
having small values of $\alpha$.  The value of $\gamma\prime$ is the slope 
of the profile at the innermost resolved point.  It is however 
an {\it apparent} rather than an 
{\it absolute} quantity (unless, of course, the inner profile does follow a 
real power-law) and as such does not reflect anything intrinsic to a 
galaxy and should therefore not be used as such.  This statement 
is of course also true when using the mean logarithmic slope $<$$\gamma$$>$ 
measured over $0.1\arcsec<R<0.5\arcsec$ (e.g., Lauer et al.\ 1995; 
Carollo \& Stiavelli 1998). 
Comparisons between the value of $\gamma\prime$ (or $<$$\gamma$$>$) for
different galaxies should be made with caution.
For example, diagrams showing these apparent quantities versus 
absolute galaxy magnitude 
are subject to the distance effects just mentioned. 

The outer power-law slope ($\beta$) of the Nuker model depends on 
how much of the 
profile's radial extent one fits; it is therefore definitely not 
a reliable parameter.  This was 
recognized from the start, and Byun et al.\ (1996) wrote:
``Even galaxies which show good agreement with the Nuker law 
within 10$\arcsec$ in general will also fail at much larger radii 
beyond the field covered by the present {\it HST} data, as the profiles 
follow a curving de Vaucouleurs law, and not a power law there.''  
It is therefore not a parameter that need be preserved in any new 
model that additionally fits the outer light-profile of early-type 
galaxies.

\subsection{Robustness of the Nuker model parameters}

Figure~\ref{fig2} shows a synthetic ``core galaxy'' profile.  
It represents a typical $r^{1/4}$ profile having an inner core. 
The structural parameters are such that 
it has an outer de Vaucouleurs profile with effective
radius $r_e = 25\arcsec$, a break radius of 0.5$\arcsec$ at $\mu=14$ 
mag arcsec$^{-2}$, and an inner power-law with slope $\gamma=0.2$. 
The radial extent which is fitted with the Nuker model is increased 
in each subsequent frame (left to right, top to bottom) in 
Figure~\ref{fig2} in order to demonstrate how the parameters of 
the fit change. 

Not surprisingly, the value of $\beta$ is strongly dependent on the
fitted radial range; this was previously known, but possibly never
quantified.  What will be surprising to many is the unstable nature of
\textit{all} the Nuker model parameters --- not just $\beta$ --- even
when fitting a (noise- and dust-free) ``core galaxy''.  As the fitted
radial extent is increased to values typically used by Rest et al.\
(2001), the Nuker model break radius marches steadily outward.  When
the mean difference between the synthetic data and the Nuker model 
reaches $\sim 0.03$ mag arcsec$^{-2}$ (the average value reported 
by the Nuker team in their fits), the derived break radius is 
twice the true break radius.  

This effect is illustrated again with two real ``core galaxy'' profiles: 
NGC~3348 from Rest et al.\ (2001) and NGC~4636 from Lauer et al.\ (1995). 
These are shown in Figures~\ref{extra1} and \ref{extra2} respectively.  
Exactly the same behavior as seen in Figure~\ref{fig2} is observed.  
It turns out that, due to the curvature in the profile beyond the 
break radius, this behavior is common to many ``core galaxies'' fitted 
with the Nuker model.  Indeed, simply by looking at the published 
``core galaxy'' profiles fitted with the Nuker model (e.g., 
Ravindranath et al.\ 2001; Laine et al.\ 2003), one can see 
for themselves how the break radii have been overestimated. 

Although 
the covariance error analysis presented for 3 galaxies in Byun et 
al.\ (1996; their Figure 6) reveals that the 10$\sigma$ $\chi^2$ 
ellipses span typically $\pm$6\% of the fitted Nuker model break 
radius, we have just witnessed that such parameter coupling is not 
the only source of uncertainty for the Nuker model parameters. 
Figure~\ref{extra1} reveals that reducing the fitted radius by 
a factor of 5 --- equivalent to imaging the same galaxy five times 
closer using the fixed Planetary Camera aperture --- can change 
$r_b$ (kpc) by a factor of 3.  These 
previously unconsidered systematic errors dominate over the random 
errors considered in Byun et al.\ (1996).  Additionally, because 
$\alpha$ couples with $\beta$ in the Nuker model, its value is 
also dependent on the radial range used.  This can in turn affect 
the value of the \textit{inner} power-law slope $\gamma$.  
As a result, the Nuker 
model's parameters are not always robust quantities: they are 
sensitive to the radial region which is fitted.

\section{The S\'ersic model}

S\'ersic's (1968) $r^{1/n}$ generalization of de Vaucouleurs' (1948) 
$r^{1/4}$ model has proved hugely successful in describing the 
light-profiles of dwarf ellipticals, ordinary ellipticals, 
and the bulges of spiral galaxies.  Early work includes that by
Davies et al.\ (1988), Capaccioli (1989), 
Caon, Capaccioli, \& D'Onofrio (1993, 1994),  
Young \& Currie (1994), James (1994), 
and Andredakis, Peletier, \& Balcells (1995).  

Recently, Graham, Trujillo \& Caon (2001b; see also Graham 2002b) showed 
a strong correlation ($r>0.8$, significance $> 99.99$\%) exists 
between the S\'ersic shape parameter $n$ and literature velocity 
dispersion measurements for those early-type galaxies studied by 
Caon et al.\ (1993) and D'Onofrio et al.\ (1994).  
Central stellar velocity dispersions are, of course, completely
independent from estimates of $n$ obtained from the galaxy 
light-profile.  This clearly shows that the S\'ersic index $n$ 
is not simply an extra parameter added to improve the fits of bulge 
light-profiles, but that it traces real physical differences in galaxies.  
A number of authors have suggested that the different profile shapes 
are connected to the gravitational potentials and total masses of 
the bulges (e.g.,  Caon et al.\ 1993; Andredakis et al.\ 1995; 
Hjorth \& Madsen 1995; Seigar \& James 1998; 
M\'arquez et al.\ 2000, 2001; Trujillo, Graham, \& Caon 2001).

The radial intensity 
distribution of the S\'ersic model is given by the expression 
\begin{equation}
I(r)=I_e \exp \left\{ -b_n \left[ \left( \frac{r}{r_e}\right)^{1/n}-1 \right] \right\},
\end{equation}
where $I_{e}$ is the intensity at the half-light radius $r_{e}$.
The quantity $b_n$ is a function of the shape parameter $n$, and is 
defined so that $r_{e}$ is the radius enclosing half 
the light of the galaxy model; it can be approximated by 
$b_n\approx 1.9992n-0.3271$, for $1\lesssim n \lesssim 10$ 
(see, e.g., Caon et al.\ 1993; Graham 2001). 
Figure~\ref{fig3} reveals the behavior of the $r^{1/n}$ model 
for values of $n$ ranging from 1 to 10; $n=4$ reproduces the 
de Vaucouleurs model, while $n=1$ reproduces an exponential profile. 

It is clear from Figure~\ref{fig3} that profiles with low values 
of $n$ (which observations show us are the low-luminosity bulges, e.g., 
Graham et al.\ 1996, their figure 11) have ``cores'' (using the definition 
$\gamma<0.3$), while profiles with large values of $n$ (the brighter 
bulges) would be described as ``power-law'' galaxies.  
However, this is exactly the opposite of what Faber et al.\ (1997) found
(e.g., their Figure~4; see also Figure~7 of Rest et al.\ 2001): cores 
are found in the \textit{brighter} bulges, while it is the relatively 
fainter bulges which have power-law centers.
To avoid confusion, it is 
important to distinguish between what might be called 
`apparent cores' from low $n$ galaxies 
and cores which have possibly been created by supermassive black 
holes in high-luminosity (high $n$) galaxies.  The former should 
perhaps not even be referred to as ``cores'' at all because they 
do not represent any departure from the inward extrapolation of the 
outer galaxy profile.  
Because there were very few low-luminosity galaxies (with probable 
S\'ersic indexes $\lesssim 3$) in the Lauer et al.\ (1995) sample, 
the ambiguity between ``apparent'' and ``real'' cores did not become 
an issue.  Studies of lower luminosity ellipticals and spiral galaxy 
bulges are however more problematic.  

Figure~\ref{fig3} also reveals that the inner profile slopes, when 
measured over the same radial range (in terms of fraction of the 
effective radius), should be equal for any sample of bulges with 
the \textit{same} S\'ersic shape.  
If, however, one used a fixed {\it angular} range (e.g., in arcseconds),  
for a sample of galaxies at a range of distances (and/or with 
intrinsically different scale-lengths), then one 
will obtain a range of different inner profile slopes, 
even if the galaxies all have the same structural shape 
(as illustrated in Section 2.1).  

For the S\'ersic model, it is simple to show that 
\begin{equation} 
\gamma\prime(r\prime) \equiv -d\log I(r\prime)/d\log r 
\end{equation}
is equal to 
\begin{equation}
(b_n/n)(r\prime/r_e)^{1/n}.\label{bunt}
\end{equation}
This can be approximated by $2(r\prime/r_e)^{1/n}$.  
Thus, at constant $(r\prime/r_e)$, $\gamma\prime$ is a monotonically 
increasing function of the S\'ersic index $n$.  Solutions to 
$\gamma\prime$ are shown in Figure~\ref{fig4}.

The value of n is well known to increase with bulge luminosity,  
and $n$ is consequently a function of position along the L-shaped 
trend seen in Figure~\ref{fig1}.  Given the correlation between $n$ and 
$\gamma\prime$ in Figure~\ref{fig4}, one would expect to see $\gamma\prime$ 
increase with bulge magnitude until a core starts to appear at the 
higher luminosity end\footnote{The S\'ersic model is known to fit 
the bulk of a bulge's light-profile, but for large values of $n$, 
its rising inner profile cannot describe the presence of cores 
in large luminous elliptical galaxies (Kormendy 1985; Lauer 1985). 
From Figure~\ref{fig1} and Figure~\ref{fig4} it is predicted that a 
better sampling of galaxies with $M_B\sim-17$ mag and S\'ersic 
indexes $1.5\lesssim n \lesssim 3$ (i.e.\ the brighter dwarf 
elliptical galaxies) should reveal more galaxies with 
$0.3< \gamma\prime <0.5$.  However, there is the issue of galaxy 
distance.}. 
%
%
This is indeed what is found in Graham \& Guzm\'an (2003; their Fig.8).

It was suggested in Section 2 that the so-called ``power-law'' 
galaxies are actually S\'ersic--$r^{1/n}$ galaxies. 
To explore how the Nuker model can imitate a pure S\'ersic profile 
when fitted to a restricted radial range, 
Figure~\ref{fig5} displays the results of fitting Nuker models to
four $r^{1/n}$ models with values of $n$ equal to 1, 2, 3, and 4. 
The fitting has been done in such a way as to try and match the 
break-radius to the radius where the change in the logarithmic profile 
slope is observed to be a maximum.  What this means is that the 
fitting routine was prevented from setting the break-radius to 
infinity\footnote{The Nuker model is equivalent to the 
$r^{1/n}$ model when $r_b\rightarrow\infty$, $\gamma=0$, and 
$\alpha=1/n$ (Byun et al.\ 1996).} (although, given the results in 
the literature, $r_b$ never tends to $\infty$  --- possibly
because of restrictions placed in the codes to keep $r_b$ bound).  
One can clearly see that 
fitting a 5-parameter function (the Nuker model) to a limited 
radial extent of a 3-parameter function (the S\'ersic model) can 
result in what many would consider a satisfactory fit. 


\subsection{Modeling the Nuker profiles with a S\'ersic model\label{fitted}}

Major-axis surface brightness profiles for 42 predominantly early-type 
galaxies imaged with the {\it HST} are given in tabular form in Lauer et 
al.\ (1995).   A logarithmically spaced sampling of the light-profiles 
was used, providing greater detail in the central regions.  These profiles 
were extracted from pre-refurbishment {\it HST} Planetary Camera images 
(taken with the F555W filter) and then 
deconvolved to account for the effects of spherical aberration 
(see Lauer et al.\ 1995 for details).  Nuker models are fitted in 
Byun et al.\ (1996)\footnote{Byun et al.\ (1996) fitted the Nuker 
model to the {\it mean} profiles, rather than the major-axis
profiles given in Lauer et al.\ (1995).}.  

Figure~\ref{fig6} shows the results of fitting the 3-parameter S\'ersic 
model to the first eight NGC galaxies in the sample of Lauer et al.\
(1995).  Figures for the remaining galaxies show the same 
behavior and are therefore not shown here.  
The data within the inner 0.13$\arcsec$, the radius inside of which 
the profiles were deemed unreliable by Lauer et al.\ (1995), 
were not included in the fitting routine.  
Quite clearly, the inner $\sim$10$\arcsec$ of some 
galaxies are very well modeled with the 3-parameter S\'ersic model.  
This is true for galaxies labeled by the Nuker team as 
either ``power-law'' galaxies (e.g., NGC~1023, NGC~1172) 
or ``core'' galaxies (e.g., NGC~720, NGC~1399).  
Relative to the S\'ersic model, NGC~1331 displays 
evidence for a large excess flux within the inner 0.3$\arcsec$.  
On the other hand, NGC~1400 appears to have a small core within 
the inner $\sim0.2\arcsec$.  

Despite the results of Figure~\ref{fig6}, we are not trying to argue 
that one should fit the inner $\sim10\arcsec$ of an appreciably 
larger profile with a S\'ersic model.  Instead, we want to 
demonstrate the inadequacy 
of the inner $\sim10\arcsec$ for drawing reliable conclusions unless 
one has further information.  For example, with only knowledge of 
the inner profile, is there really an evacuated core in NGC~720?  
Fitting a S\'ersic model one would say no, fitting a Nuker model 
one would say yes. 


In the following section we will advocate 
a new definition for a core, specifically, a deficit of central 
starlight (not due to dust) relative to the inward extrapolation of 
the outer light-profile.  
This differs from the Nuker team's definition which is dependent 
on the inner profile slope. 


\section{A new empirical light-profile model}

We have discussed how fitting a Nuker model could lead to a false conclusion 
regarding the existence of a (partially evacuated) core.  Additionally, 
we have just seen in Figure~\ref{fig6} that if one only has the inner 
portion of a larger 
profile, then a S\'ersic model is also capable of fitting the data --- 
even when a real core may be present. 
One obvious way to avoid this potential confusion, and at the same time
enable one to connect the inner and outer galaxy structure, 
is to increase the radial extent of the galaxy's surface 
brightness profile one is investigating.  Given that galaxies 
with evacuated cores probably do exist, we need 
a new model to describe the entire radial extent of a profile.  
Of course, the additional data points in the outer profile provide more 
information than contained in the inner few arcseconds and therefore 
enable one to 
determine an additional parameter beyond the capabilities of the
Nuker model.  Specifically, one can fit for the
curvature in the outer profile where a power-law is known
to be inadequate for the majority of galaxies.  

Modifying the S\'ersic model through the inclusion of an inner 
power-law, or similarly, modifying the Nuker model through the 
transformation of the outer power-law to a S\'ersic function, 
one obtains the expression 
\begin{equation} 
I(r)=I' \left[ 1+\left( \frac{r_b}{r}\right)^{\alpha} \right]^{\gamma/\alpha}
\exp \left\{ -b_n[(r^{\alpha}+r_b^{\alpha})/r_e^{\alpha}]^{1/(\alpha n)}\right\}, 
\label{bomba}
\end{equation}
where $r_b$ is the break-radius separating the inner power-law having 
logarithmic slope $\gamma$ from the outer S\'ersic function having a 
shape parameter $n$ and effective half-light radius $r_e$. 
The quantity $b_n$ is a function of $n$ and has the usual meaning 
(see section 3).  By leaving $b_n$ defined this way, 
the value of $r_e$ is the effective half-light radius of
the outer $r^{1/n}$ profile beyond the transition region, 
and not the half-light radius of the new model.
The effective surface brightness of the outer S\'ersic profile 
is obtained by putting $r=r_e$ and $r_b=0$ in equation~\ref{bomba}, 
while $r_b$ retains its value in equation~\ref{iprime} below.  
The intensity $I_b$ at the break-radius $r_b$ can be evaluated 
from the expression 
\begin{equation}
I' = I_b 2^{-(\gamma/\alpha)} \exp \left[ b_n(2^{1/\alpha}r_b/r_e)^{1/n}\right]. 
\label{iprime}
\end{equation}
The final parameter, $\alpha$, controls the sharpness of the transition
between the inner (power-law) and outer (S\'ersic) regimes --- higher values
of $\alpha$ indicating sharper transitions.  It can be held fixed (for
example, $\alpha = 100$ is a good approximation to a perfectly sharp 
transition), or it can be varied if one is interested in accurately 
matching the transition region.
In practice (see Trujillo et al.\ 2003), we find that a sharp transition 
($\alpha \gtrsim 3$) may be preferable; high values of $\alpha$ also 
minimize possible coupling of the parameters, which may compromise the 
inner profile slope (as can happen with Nuker model).  

In spite of the previously discussed issues associated with 
the inner power-law slope of the Nuker model (section 2.1), 
$\gamma$ has remained a quantity of interest in the literature. 
Even with a sharp transition (i.e.\ large values of $\alpha$) 
our new empirical model has curvature built into it through the 
parameter $n$.  Consequently, the problems that the Nuker
model had in trying to accommodate curvature (via small values of 
$\alpha$) should not plague this new model, and values 
of $\gamma$ derived from fitting equation~\ref{bomba} should more 
accurately reflect the observed inner profile slope.  

Despite appearances, equation~\ref{bomba} is remarkably simple.
When $r$ is less than $r_b$ equation~\ref{bomba} approximates a simple
power-law with slope $\gamma$; when $r$ is greater than $r_b$ 
equation~\ref{bomba} represents a S\'ersic model.   
There is however a variable transition region, as mentioned above, 
which depends on the value of $\alpha$. 
Setting $r_b$ and $\gamma$ equal to zero, one recovers a pure 
$r^{1/n}$ model at all radii. 
Additionally, when $n\rightarrow \infty$ 
it can be shown that 
equation~\ref{bomba} approximates two power-laws separated at $r_b$.
Figure~\ref{fig7} shows equation~\ref{bomba} for different sets of 
parameters for this new model. 

Figure~\ref{fig8} presents two examples where the new empirical 
model has been applied to real galaxies.  One can see that the 
``power-law'' galaxy NGC~5831 from Rest et al.\ (2001) is well 
described by a S\'ersic model over its entire observed radial 
range (0.1$\arcsec$ to $\sim 3 r_e$).  On the other hand, 
NGC~3348 from Rest et al.\ (2001) clearly displays a flattening of the 
inner profile relative to the outer S\'ersic profile; this break 
is well matched by the new model.\footnote{NGC~4636 (Fig.~\ref{extra2}) 
is not shown because given its size of 6$\times$4.7 arcminutes, 
we have no outer light-profile in the {\it HST} image.}  
In Trujillo et al.\ (2003), we present these and similar fits to 
approximately twenty bona fide elliptical galaxies.  Using 
archival {\it HST} data matching that of the inner profiles from 
Lauer et al.\ (1995) or Rest et al.\ (2001), we verify that ``power-law''
galaxies have pure S\'ersic profiles, while core galaxies
are well fit by our new model. 

We have explored the stability of the parameters in the new model 
and found that in order to obtain a robust estimate of the  
curvature in the outer profile, one requires a profile which 
extends to, typically, at least the half-light galaxy radius.  
Although, we would advocate that one should fit as much of the 
profile as possible, preferably out to 2--3 $r_e$.  As our 
main drive has been the issue of recovering core sizes, we 
remark here on our fits to the ``core'' galaxy NGC~3348.  Fitting 
equation~\ref{bomba} to our full 70$\arcsec$ profile (Figure~\ref{fig8}), 
we obtained a break radius of 0.45$\arcsec$, 
$\mu_b=15.31$ mag arcsec$^{-2}$, $\gamma=0.18$, $r_e$=21.82$\arcsec$, 
and $n=3.87$.  Truncating the profile one data point at a time, these
numbers changed by at most 5\%.  Upon reaching a profile with a radial 
extent of 21.5$\arcsec$ ($\sim 1 r_e$), application of 
equation~\ref{bomba} gave $r_b=0.45\arcsec$, $\mu_b=15.31$ mag 
arcsec$^{-2}$, $\gamma=0.18$, $r_e=20.84\arcsec$, and $n=3.81$, 
with an r.m.s.\ scatter of 0.028 mag for the fit.  Thus, use of the 
complete 
galaxy light-profile (or at least enough of the profile to reliably 
determine the curvature beyond the break radius) can enable one to 
make robust estimates of the core size.  Doing this, the core size we 
obtained for NGC~3348 is more than a factor of two smaller than the 
value published in Rest et al.\ (2001).  The reason for this is 
apparent in Figures~\ref{fig2} -- \ref{extra2}. 
Fitting the full 55$\arcsec$ profile of NGC~5831 yielded 
$r_e=28.28\arcsec$ and $n=4.91$; fitting to only 27.3$\arcsec$ 
($\sim 1 r_e$) gave $r_e=27.00$ and $n=4.83$, within 
5\% of each other. 

Lauer et al.\ defined a ``core'' to be ``the region interior to a 
sharp turndown or break in the steep outer brightness profile, 
provided that the profile interior to the break has $\gamma<0.3$.''
We suggest an alternative definition, such that a core 
refers to a deficit (not due to dust) of flux relative to the inward 
extrapolation of the outer S\'ersic profile.  
A ``core'' would thus refer to something which is likely to be 
real, rather than an apparent feature which can appear in plots of 
$\mu$ versus $\log r$ (see Figure~\ref{fig5}).

Consideration of this new model (equation~\ref{bomba}) is also suggested 
for studies requiring a realistic gravitational lens model for elliptical
galaxies at intermediate redshifts 
(e.g., Evans \& Wilkinson 1998; Mu\~noz, Kochanek, \& Keeton 2001; 
Chae 2002; Keeton 2002).  Other 
interesting areas of research are the feeding of SMBHs via the capture of 
stars and dark matter (e.g.\ Zhao, Haehnelt, \& Rees 2002) and the 
evaluation of the central 
mass deficit possibly excavated by coalescing massive black holes 
(Ebisuzaki et al.\ 1991; Faber et al.\ 1997; 
Milosavljevic \& Merritt 2001; Milosavljevic et al.\ 2002:
Ravindranath, Ho, \& Filippenko 2002; Komossa et al.\ 2003). 
Current research assumes that 
galaxies initially had the same inner profile\footnote{Current
assumptions that faint elliptical galaxies (the
assumed building blocks of core galaxies) have isothermal cusps 
with $\rho\sim r^{-2}$ may not be correct.  Low-luminosity
ellipticals have small values of $n$ and therefore shallow inner
cusps (Figure~\ref{fig3}, see also Graham \& Guzm\'an 2003), 
although they may possess additional central components.} 
(such as an $r^{1/4}$ profile or a density profile where 
$\rho(r)\sim r^{-2}$, e.g. Volonteri, Haardt, \& Madau 2003) 
and compares this with how the profile actually looks. 
It might be of interest to replace this assumption with the 
inward extrapolation of the observed outer S\'ersic profile.  
Bulges {\it are not} structurally homologous systems; their properties
vary systematically with total luminosity and mass.  To assume 
structural homology very likely introduces a systematic bias into 
these types of analysis. 
Lastly, Ravindranath et al.\ (2002) have used the break radii
from the Nuker model to estimate the evacuated core masses 
in their sample of galaxies (Ravindranath et al.\ 2001) 
and those from Rest et al.\ (2001).  This may explain why 
they found a weaker trend than expected, and one with considerable 
scatter, between the central SMBH mass and the ejected mass. 
This issue will be explored in Trujillo et al.\ (2003, in prep). 
%

\section{Summary}

None of the 5 parameters of the Nuker model are found to be robust.
Hence, they can not represent any fixed physical quantity.  Recent methods 
which have tried to circumvent one of the Nuker parameters by measuring 
the inner power-law slope at some fixed radius in arcseconds 
($\gamma\prime$), but without taking galaxy distance or size into 
account, are shown (quantitatively) to be subject to strong biases.
Measured values of $\gamma\prime$ (and $<$$\gamma$$>$) are not 
intrinsic to a galaxy, and 
can change considerably if the same galaxies are located at different 
distances without any actual change to the intrinsic galaxy structure. 

\begin{figure}
\epsscale{0.75}
\plotone{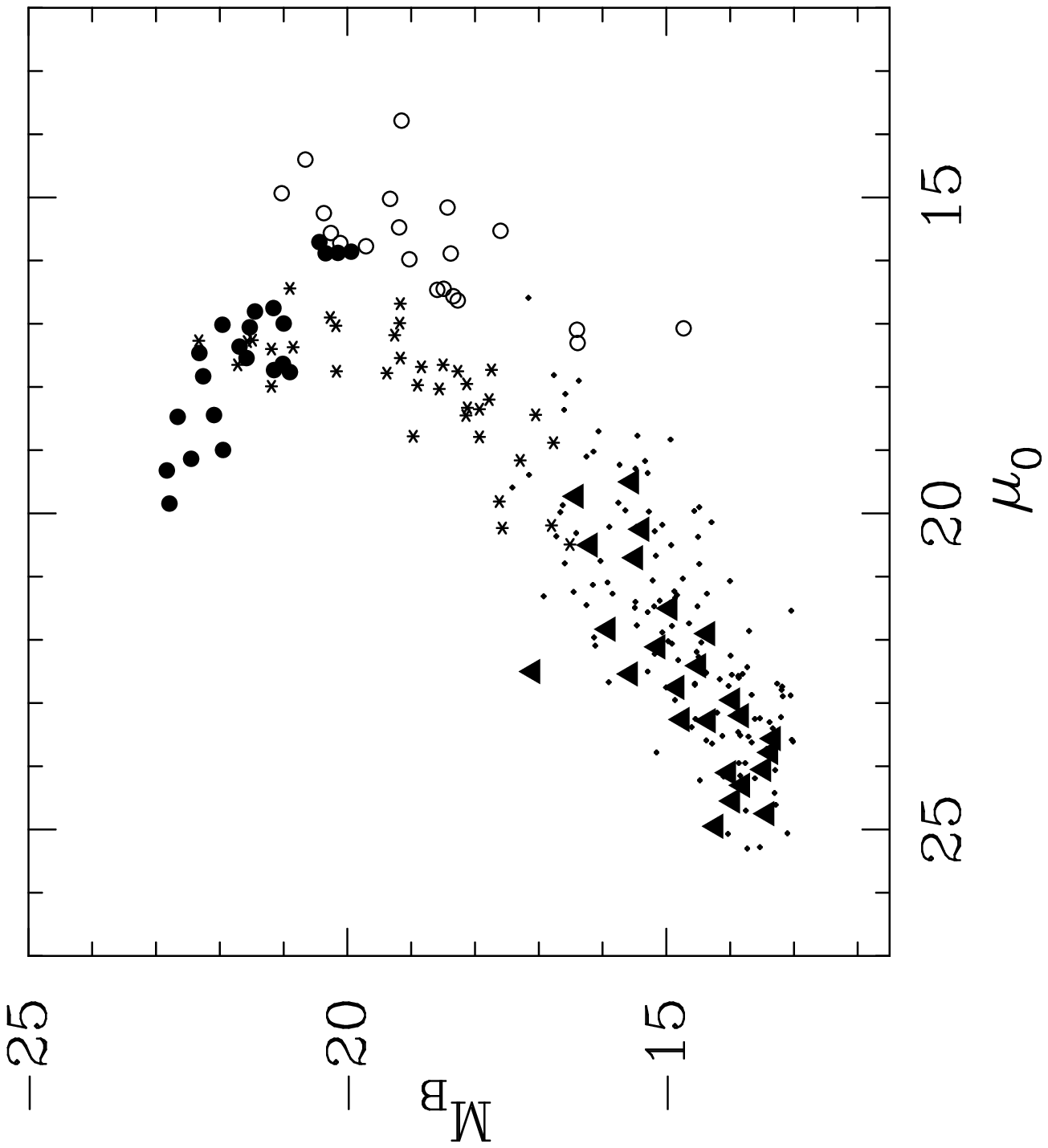}
\caption{Absolute $B$-band galaxy magnitude $M_B$ versus central $B$-band surface 
brightness $\mu_0$ for the entire gamut of elliptical galaxies.  The dwarf elliptical 
galaxies from Binggeli \& Jerjen (1998; see also Jerjen \& Binggeli 1997) 
are points, while the dwarf elliptical galaxies from Stiavelli et al.\ (2001) 
are triangles.  The intermediate luminosity 
Virgo and Fornax elliptical galaxies from Caon et al.\ (1993) and
D'Onofrio et al.\ (1994) are given as stars, the so-called ``power-law'' 
galaxies from Faber et al.\ (1997) are open circles, while their ``core''
galaxies are filled circles.  
The outlying dwarf elliptical galaxy from Stiavelli et al.\ (2001) is VCC9, and the 
outlying ``power-law'' galaxy is V1627. 
$H_0=70$ km s$^{-1}$ Mpc$^{-1}$ was used, as was a Virgo distance modulus of 31.2. 
}
\label{fig1}
\end{figure}

\begin{figure}
\epsscale{0.75}
\plotone{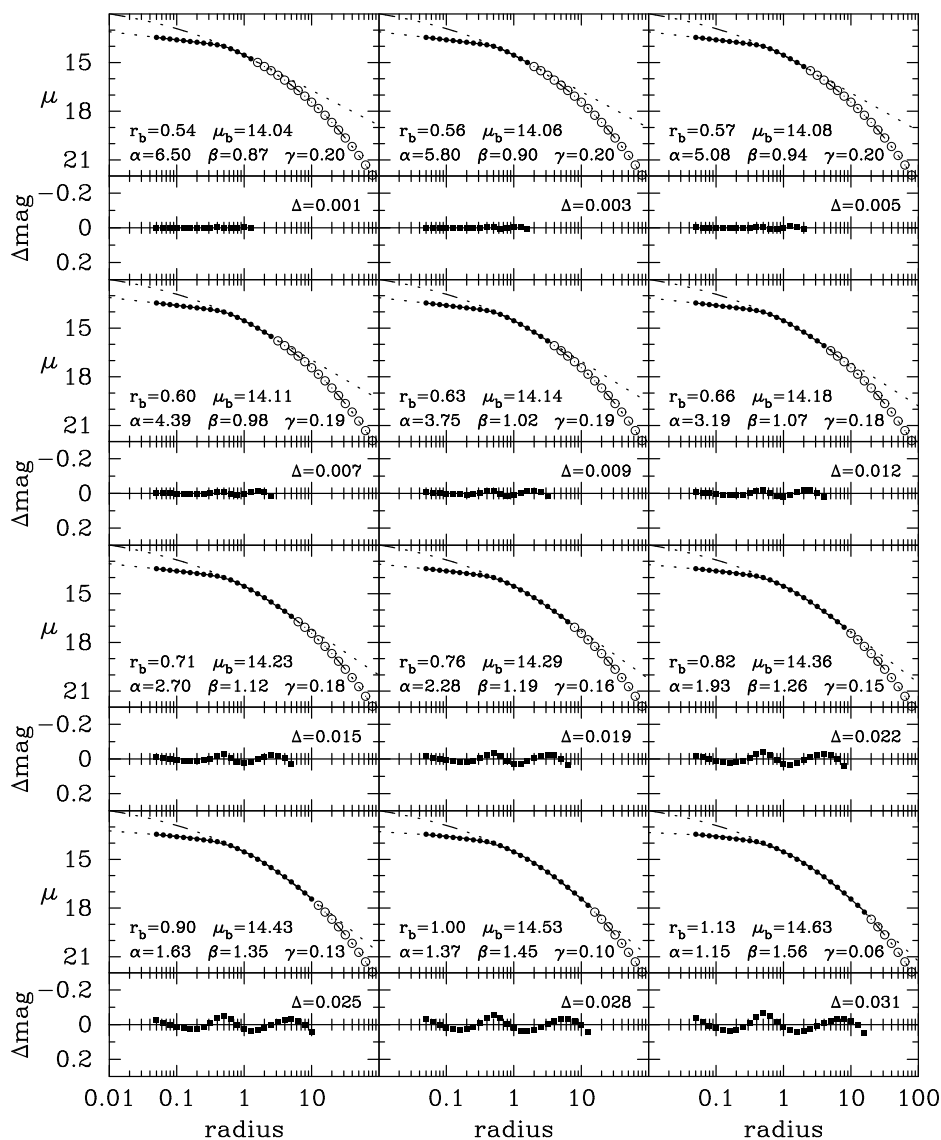}
\caption{A model ``core'' galaxy, along with Nuker-model fits
using progressively larger radial ranges.    
The best-fitting Nuker models (dotted lines) have been derived 
from modeling a profile consisting of an outer de Vaucouleurs 
profile (dash-dot-dot-dot curve) having $r_e=25\arcsec$ 
and an inner power-law with a negative logarithmic slope of 0.2.  The 
break radius of this model, delineated by the filled and open circles, 
is at 0.5$\arcsec$ and $\mu_b$=14.0 mag arcsec$^{-2}$, and a value of 
$\alpha=8.0$ (see equation 5) has been used.  
The radial extent of the fitted data (filled circles) is 
increased from left to right and top to bottom.  Although every fit looks
acceptable, the actual Nuker model parameters (inset in figure) can be 
seen to vary systematically.  The r.m.s.\ scatter $\Delta$ mag for 
each fit is given in each panel.  
}
\label{fig2}
\end{figure}

\begin{figure}
\epsscale{0.75}
\plotone{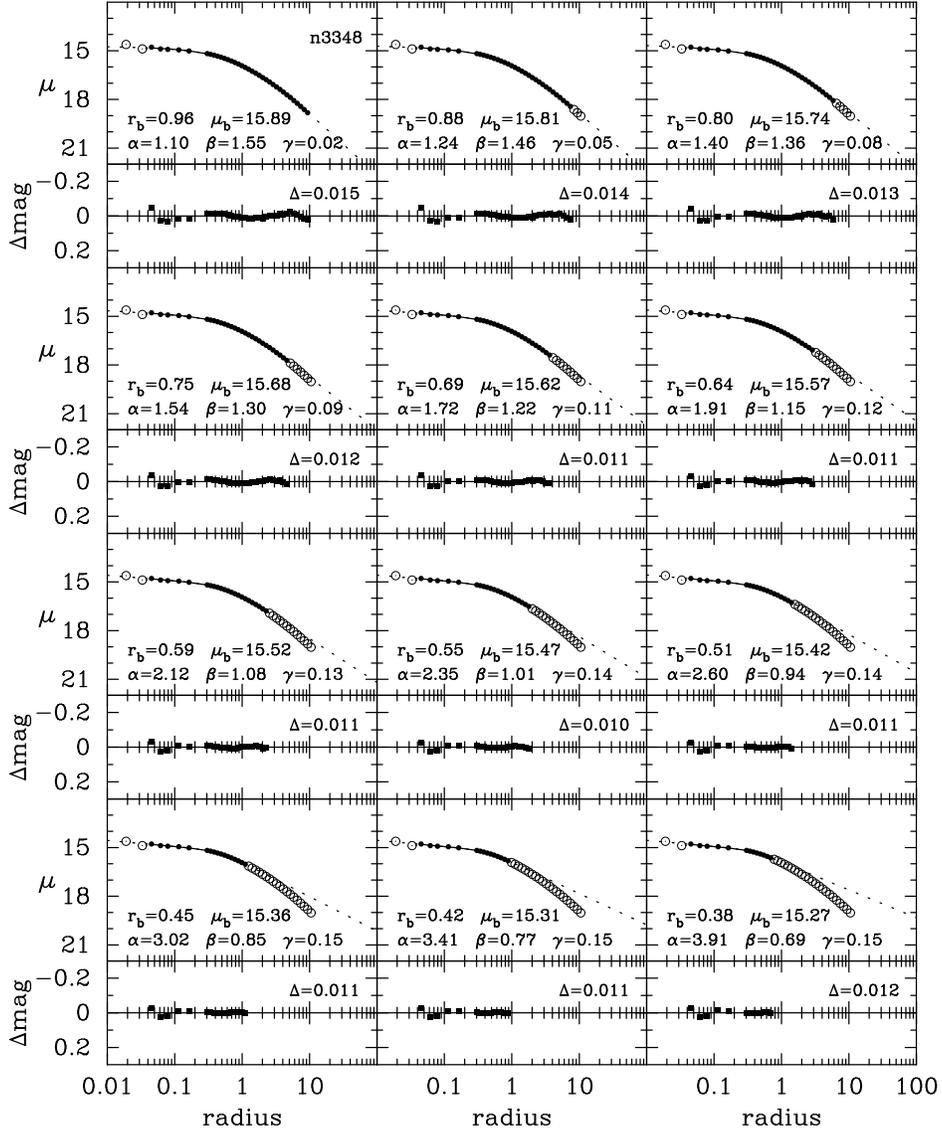}
\caption{
Nuker-model fits (solid and dotted lines), using
progressively smaller radial ranges, are applied to the major-axis
F702W surface-brightness profile of NGC~3348 (from Rest et al.\ 2001).
The radial extent of the fitted data (filled circles) decreases from
left to right and top to bottom.  Although every fit looks acceptable,
as we saw in figure~\ref{fig2} the actual Nuker model parameters (inset in
figure) vary systematically with the fitted radial extent (c.f\ Fig.~\ref{fig2}.  
Rest et al.\ (2001) reported a break radius $r_b = 0.99\arcsec$ for this
galaxy, based on their Nuker-model fit. 
} 
\label{extra1}
\end{figure}

\begin{figure}
\epsscale{0.75}
\plotone{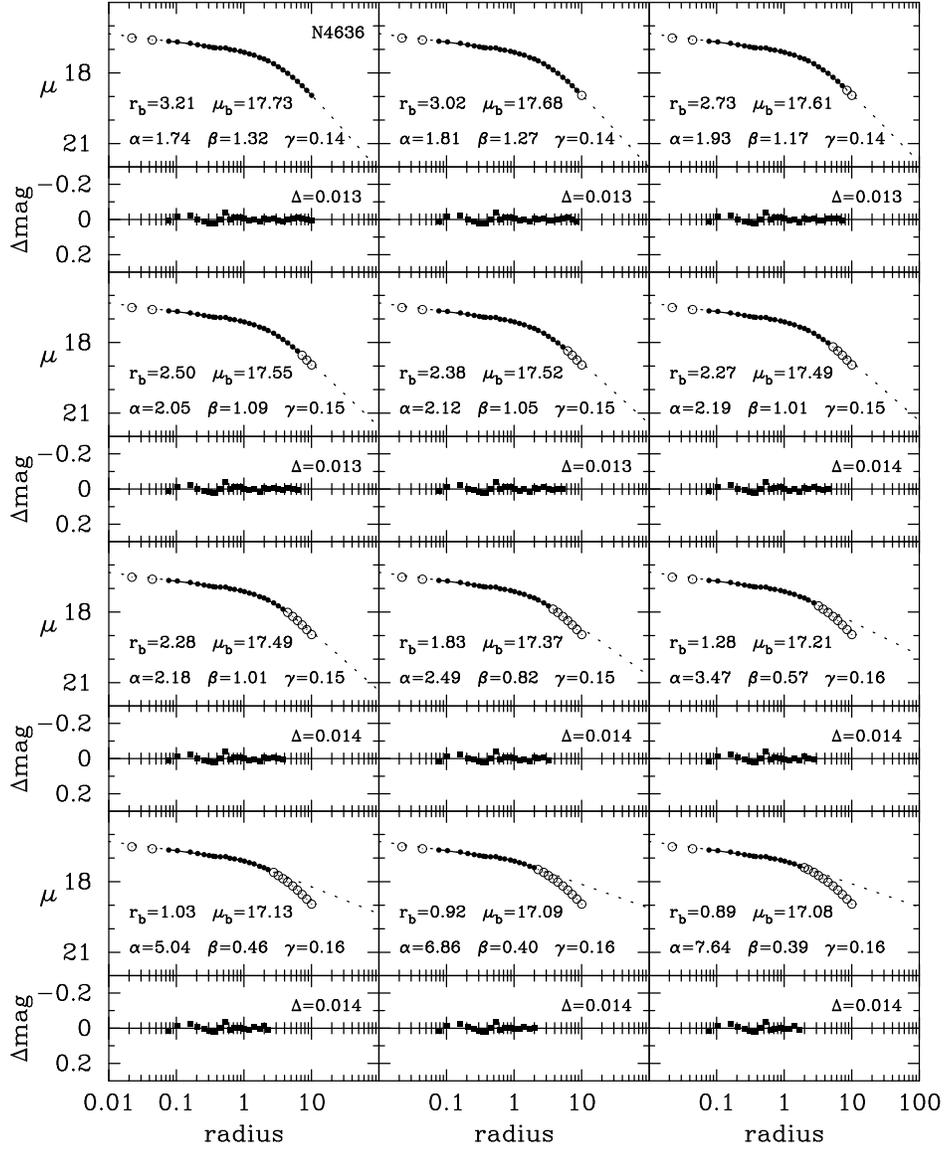}
\caption{
Same as Figure~\ref{extra1}, but using the mean-axis F555W
galaxy profile of NGC~4636 from Lauer et al.\ (1995).
Byun et al.\ (1996) reported a break radius of 3.21$\arcsec$ 
for this galaxy.  If this galaxy had of been five times closer, 
a radius five times smaller (in physical units) would have been 
sampled with a 10$\arcsec$ profile, and a break radii some 3 times 
smaller may have been reported.  
}
\label{extra2}
\end{figure}

\begin{figure}
\epsscale{0.75}
\plotone{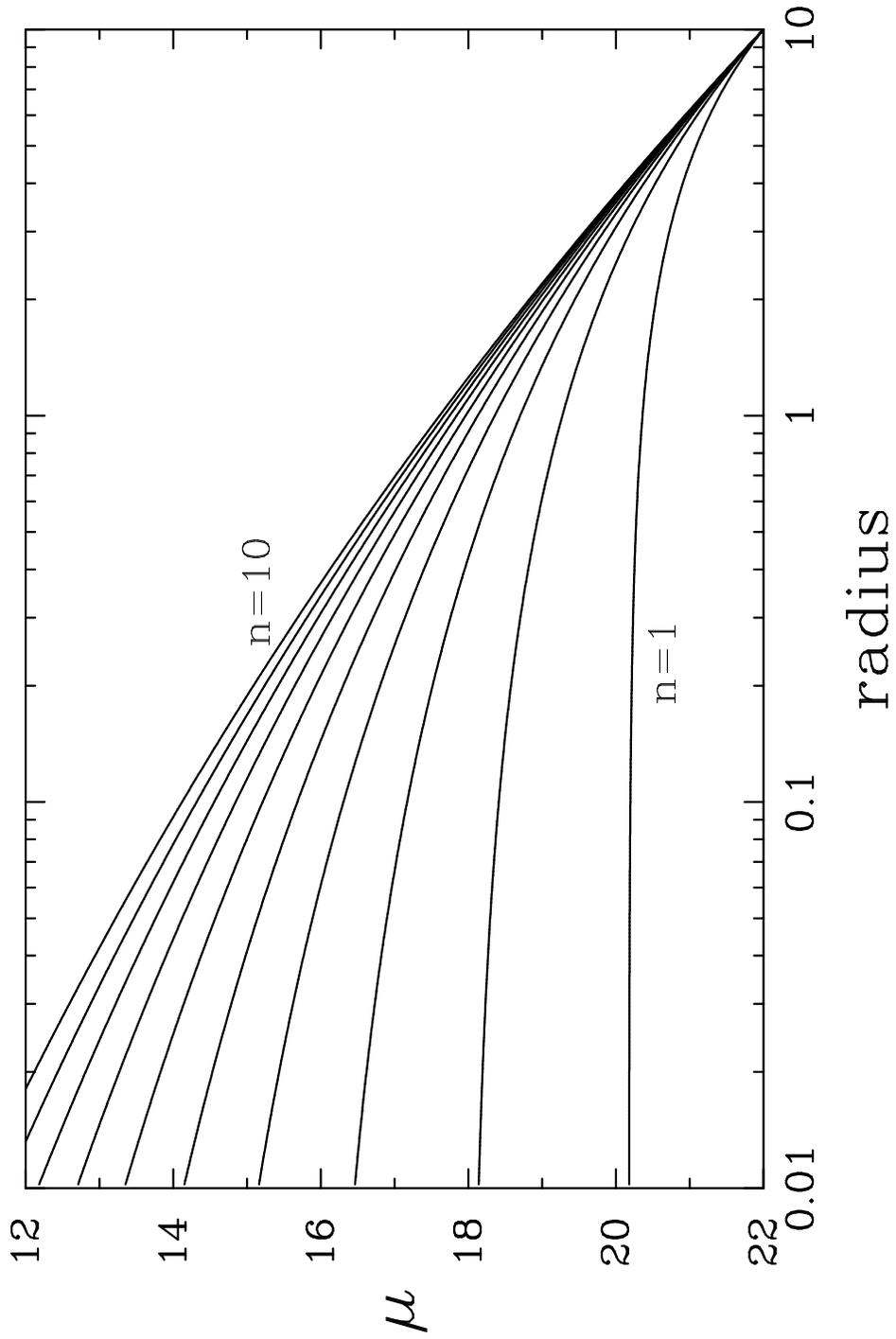}
\caption{
S\'ersic profiles with values of $n$ ranging from 1 to 10 
are shown on a logarithmic scale.  The effective half-light
radius ($r_e$) is equal to 10 for each profile. 
}
\label{fig3}
\end{figure}

\begin{figure}
\epsscale{0.75}
\plotone{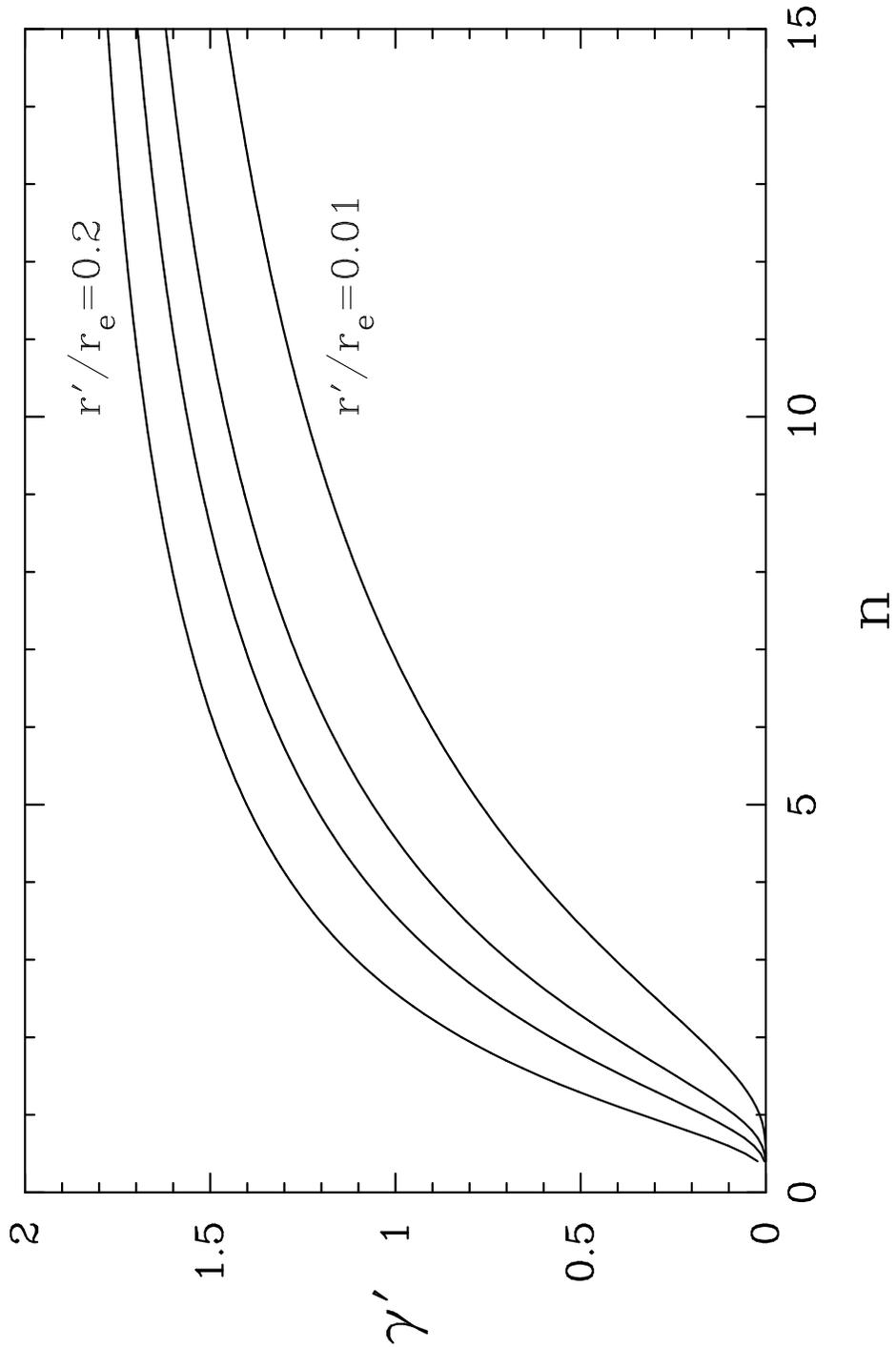}
\caption{
The negative logarithmic slope of the S\'ersic model ($\gamma\prime$), 
at different fractions of the half-light radius $r_e$ 
(0.01, 0.05, 0.1, 0.2), is shown as a function of the S\'ersic index $n$. 
}
\label{fig4}
\end{figure}

\begin{figure}
\epsscale{0.75}
\plotone{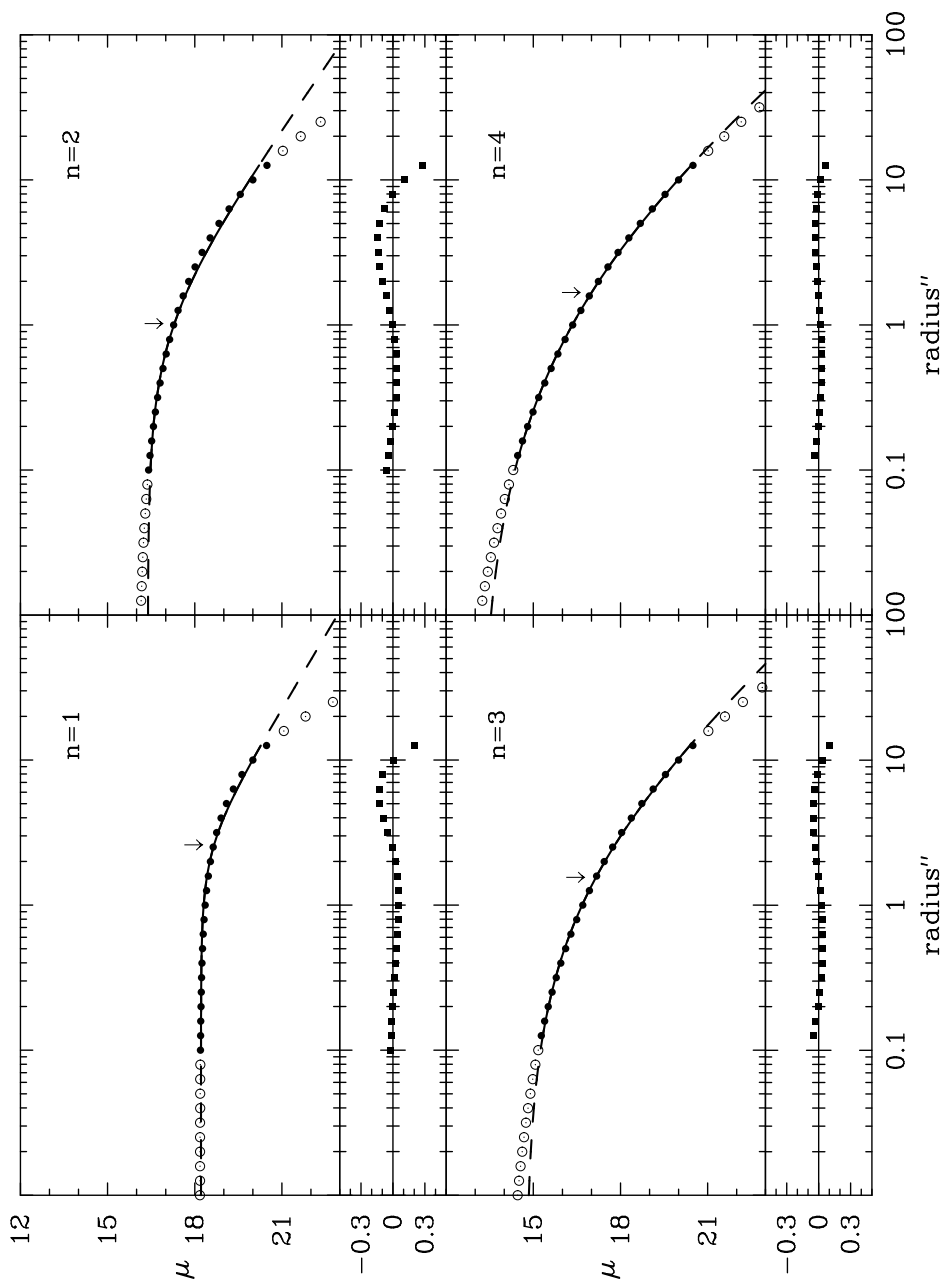}
\caption{
Four S\'ersic profiles (small circles) with $n$=1,2,3, and 4, 
$r_e=10\arcsec$, and $\mu_e=20$ mag arcsec$^{-2}$.  
The solid lines are Nuker model fits to the data points (filled circles); 
the dashed lines are extrapolations outside the region of the fit. 
The Nuker model break-radii are indicated with arrows. 
}
\label{fig5}
\end{figure}

\begin{figure}
\plotone{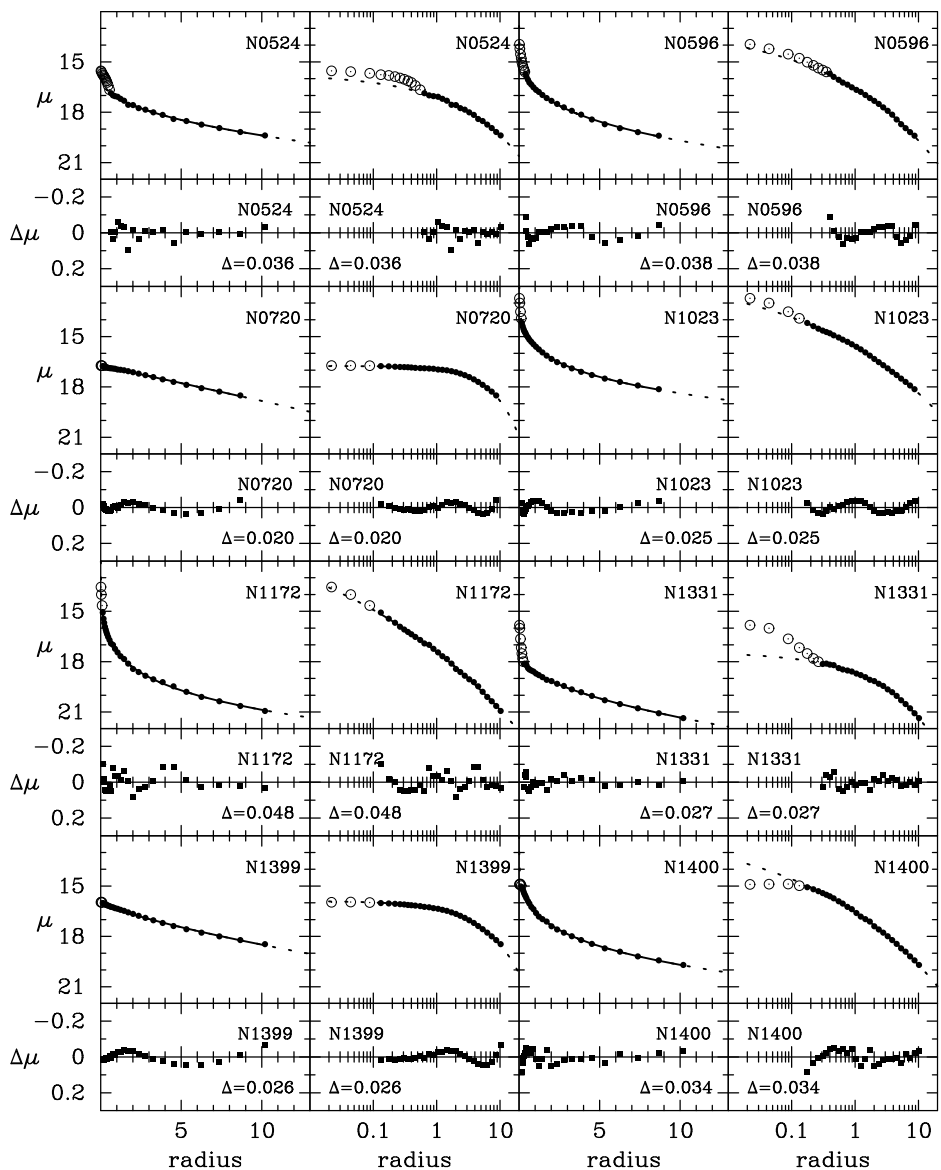}
\caption{
S\'ersic fits to the first eight NGC galaxy profiles 
published in Lauer et al.\ (1995).  
Dots indicate data points used in the fit, while 
open circles indicate points not used for the fit
(c.f.\ Byun et al.\ 1996, their Fig.3). 
The radius is in arcseconds. 
}
\label{fig6}
\end{figure}

\begin{figure}
\epsscale{0.75}
\plotone{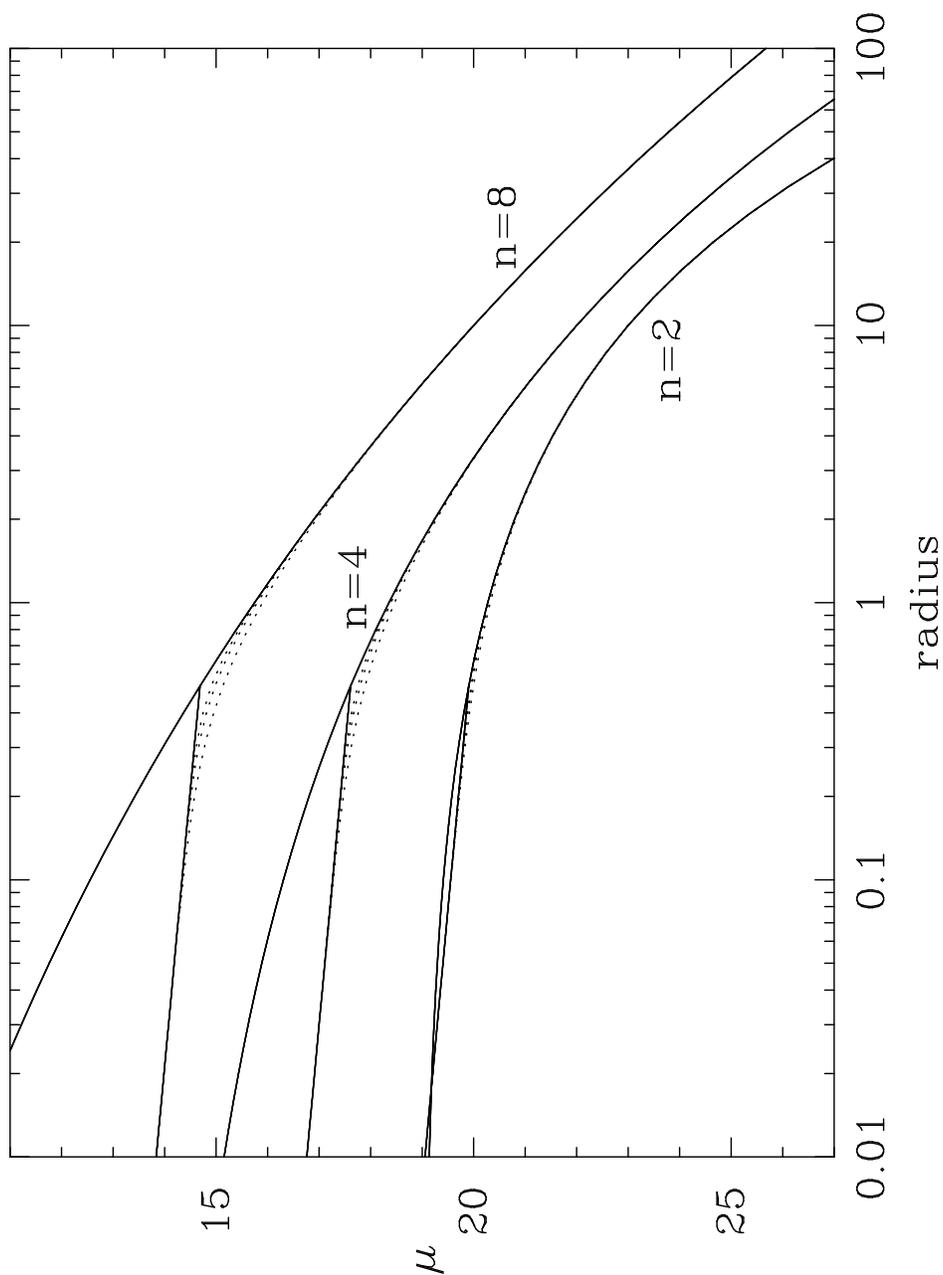}
\caption{
The new empirical model given in Equation~\ref{bomba} is illustrated 
by the dotted curves for a range of structural parameters.  Profiles 
with values of $\alpha$ equal to 2, 3, and 4 are shown, the latter 
giving the sharpest transition.  In all models $r_e=10$, $r_b=0.5$, 
and $\gamma=0.2$.
For comparison, an inner power-law with slope equal to -0.2 is shown (solid 
line) as are S\'ersic profiles (solid curves) having the same S\'ersic 
shape index $n$ as the new empirical model (values of $n$ are given in 
the Figure).  
}
\label{fig7}
\end{figure}

\begin{figure}
\epsscale{0.75}
\plotone{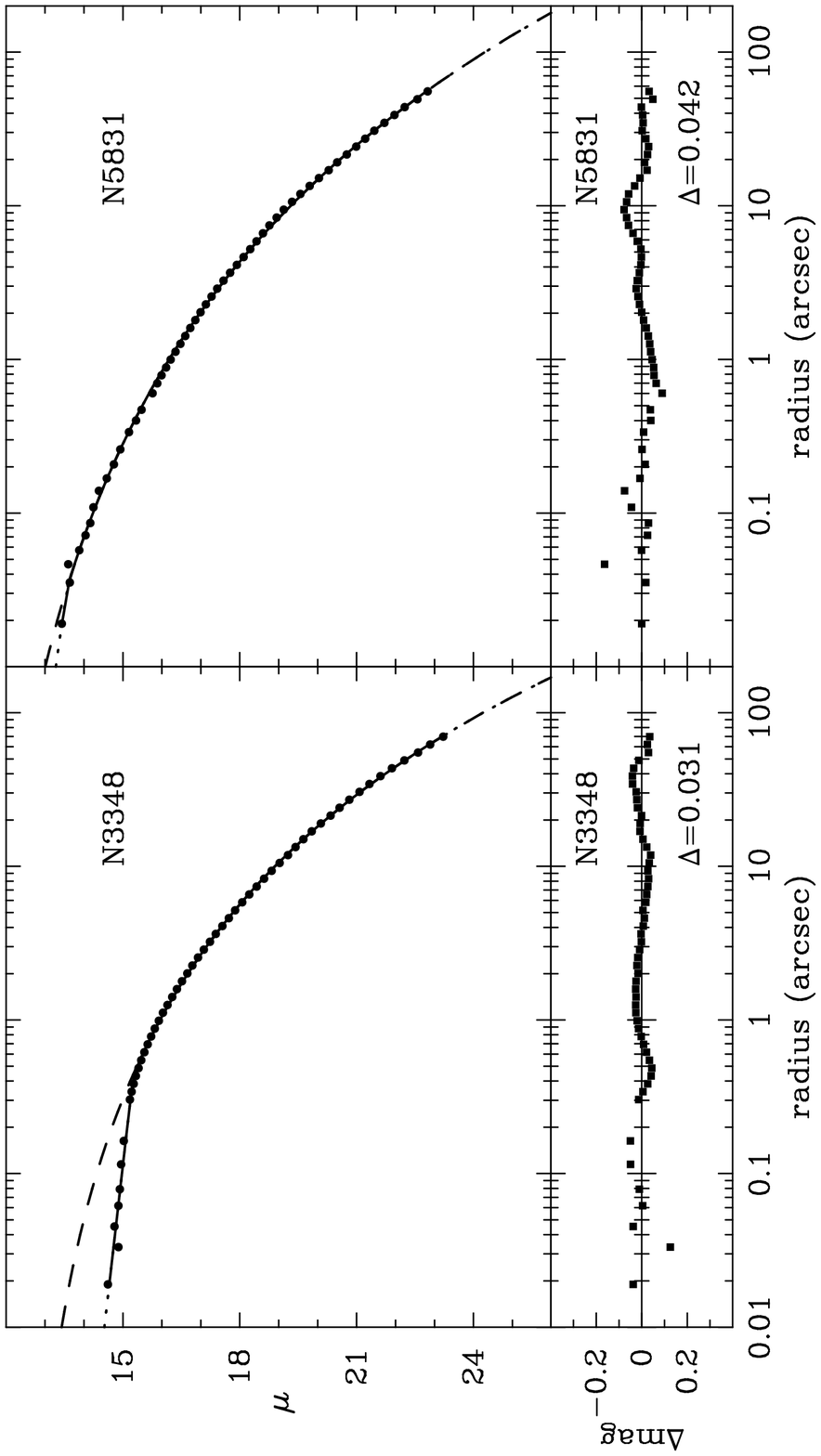}
\caption{
Two {\it HST} major-axis surface-brightness profiles of elliptical galaxies.  The
inner $r \lesssim 15\arcsec$ are the (deconvolved) profiles from Rest et
al.\ (2001), while the outer points are from the full WFPC2 mosaic of the
same exposures (see Trujillo et al.\ 2003 for full details).
The solid lines are fits using the 
new empirical model of equation~\ref{bomba}, with inner and outer extrapolations
indicated by the dotted lines.  Extrapolations of the outer, S\'ersic-like
part of the model inward, past the break radius, are indicated by the
dashed lines.  
For NGC~5831, a ``power-law'' galaxy according to Rest et al.\ (2001) with 
$r_b = 1.78\arcsec$, our best fit is essentially a pure S\'ersic model.  
In NGC~3348, which Rest et al.\ (2001) classified as a ``core'' galaxy with 
break radius $r_b = 0.99\arcsec$, there is a clear inner break from
the S\'ersic profile; we find $r_b = 0.45\arcsec$.  
The r.m.s.\ scatter $\Delta$ mag for the fit is given in the plot.  
}
\label{fig8}
\end{figure}

As observed by previous authors (e.g.\ Caldwell 1983; 
Jerjen \& Binggeli 1997) dwarf 
elliptical galaxies form a continuous extension to the intermediate 
luminosity elliptical galaxies in the central surface brightness --- 
absolute magnitude plane (Figure~\ref{fig1}), 
such that $M_B \propto (3/5)\mu_{0,B}$.  
Faber et al.\ (1997) wrote, ``A major conclusion
is that small hot galaxies are much denser than large ones'' and 
that ``the apparent turndown in [central] surface brightness at faint 
magnitudes ...\ is probably a resolution effect''.  
However, observations of the {\it underlying host galaxy} (i.e., 
excluding nuclear sources) reveal, 
over a range of 7.5 mag, that small hot galaxies are actually less 
centrally dense than larger hot galaxies - this is not an artifact 
of resolution.  We instead attribute the observed turndown in 
central surface brightness at {\it bright} magnitudes ($M_B \lesssim -20.5$ mag) 
to the presence of galactic cores.  
Lastly, it is noted that the lower luminosity elliptical galaxies 
are known 
to be well described by the S\'ersic model, and it is suggested here 
that the ``power-law'' galaxies 
may in fact simply be S\'ersic--$r^{1/n}$ galaxies with no resolved core. 
%
The 3-parameter S\'ersic model provides a remarkably 
good fit to the 42 inner galaxy light-profiles initially studied by the Nuker team. 
In some cases the quality of the fit may be because the so-called 
``power-law''  
galaxies are S\'ersic--$r^{1/n}$ galaxies all the way into the resolution 
limit, while in other cases it is probably a result of the limited radial 
extent of the profiles. 
%
%

Whether or not a ``core'' represents a real physical change, 
or just an apparent change in profile slope, can be determined by looking at 
the entire light-profile, rather than just the central profile, and we 
advocate a new definition for a ``core'' as a deficit in central flux
relative to the outer S\'ersic profile.  
The results of doing this are shown in Trujillo et al.\ (2003) for 
a sample of bonafide elliptical galaxies.  In order to model the entire
light-profiles, a new empirical model comprised of an outer S\'ersic function 
and an inner power-law has been developed and is presented here
and will be described in greater mathematical detail in Trujillo 
et al.\ (2003). 

By combining an inner power-law with an outer S\'ersic 
profile, one should be better able to: 

\noindent
1. Explore where and how the $r^{1/n}$ model fails to provide a good match 
to the inner light-profile, and thereby 
test whether the so-called ``power-law'' galaxies are actually 
galaxies described by an $r^{1/n}$ model down to the 
resolution limit (i.e.\ having no resolvable cores, and not having an inner 
power-law profile).  

\noindent
2. Quantify central excess fluxes known to exist in many galaxies.

\noindent
3. Search for connections between the shape of a galaxy's outer profile, 
as represented by $n$, and the properties of its core.  

\noindent
4. Quantify the slope and break-radii of ``cores'' normalized to
the galaxy's effective half-light radius. 

\noindent
5. Model the gravitational lensing deflection caused by distant 
elliptical galaxies. 

\noindent
6. Search for correlations between supermassive black hole mass
(possibly derived from the $\log n$--$\log M_{\rm bh}$ relation, 
Graham et al.\ 2001a, 2003, Erwin et al.\ 2003), 
break-radii, and the central flux/mass deficit in galaxies having  
partially evacuated cores. 

\acknowledgements
We wish to thank Nicola Caon for supplying us with the observed
central surface brightness values for the Virgo and Fornax elliptical 
galaxies shown in Figure~\ref{fig1}.  We are also happy to thank Vicki 
Sarajedini for kindly proofreading this work. 
This research was supported in part by NASA through the American 
Astronomical Society's Small Research Grant Program.

\clearpage

\newpage
\begin{references}
\reference{APB95} Andredakis, Y.C., Peletier, R.F., \& Balcells, M.\ 1995, \mnras, 275, 874 
\reference{AaL01}Alexander, T., \& Livio, M.\ 2001, ApJ, 560, L143
\reference{BaJ98}Binggeli, B., \& Jerjen, H.\ 1998, A\&A, 333, 17
\reference{Bet96}Byun, Y.-I., et al.\ 1996, AJ, 111, 1889 
\reference{Cal83}Caldwell, N.\ 1983, AJ, 88, 804
\reference{CCD93}Caon, N., Capaccioli, M., \& D'Onofrio, M.\ 1993, \mnras, 265, 1013
\reference{CCD94}Caon, N., Capaccioli, M., \& D'Onofrio, M.\ 1994, A\&AS, 106, 199
\reference{Cap89}Capaccioli, M.\ 1989, in The world of galaxies, ed. H. G. Corwin, L. Bottinelli (Berlin: Springer-Verlag), 208
\reference{CaS98}Carollo, C.M, \& Stiavelli, M.\ 1998, AJ, 115, 2306
\reference{Cha02}Chae, K.H.\ 2002, ApJ, 568, 500
\reference{CGJ02}Choi, P.I., Guhathakurta, P., \& Johnston, K.V.\ 2002, AJ, 124, 310
\reference{Cet93}Crane, P., et al.\ 1993, AJ, 106, 1371
\reference{Det88}Davies, J.I., Phillips, S., Cawson, M.G.M., Disney, M.J., \& Kibblewhite, E.J.\ 1988, MNRAS, 232, 239
\reference{deV48}de Vaucouleurs, G.\ 1948, Ann.\ d'Astrophys., 11, 247
\reference{DOn01}D'Onofrio, M.\ 2001, MNRAS, 326, 1517
\reference{DCC94}D'Onofrio, M., Capaccioli, M., \& Caon, N.\ 1994, \mnras, 271, 523
\reference{ETO91}Ebisuzaki, T., Makino, J., \& Okumura, S.K.\ 1991, Nature, 354, 212 
\reference{EGC03}Erwin, P., Graham, A.W., \& Caon, N.\ 2003, in Carnegie Observatories Astrophysics Series, Vol. 1: Coevolution of Black Holes and Galaxies, ed. L.C.\ Ho (Pasadena: Carnegie Observatories, http://www.ociw.edu/ociw/symposia/series/symposium1/proceedings.html)
\reference{EaW98}Evans, N.W., \& Wilkinson, M.I.\ 1998, MNRAS, 296, 800
\reference{Fet87}Faber, S.M., Dressler, A., Davies, R.L., Burstein, D., Lynden-Bell, D.\ 1987, in Nearly Normal Galaxies: From the Planck Time to the Present, edited by S.M.\ Faber (Springer, New York), p.\ 175
\reference{Fet97}Faber, S.M., et al.\ 1997, AJ, 114, 1771 
\reference{Fet94}Ferrarese, L., van den Bosch, F.C., Ford, H.C., Jaffe, W., \& O'Connell, R.W.\ 1994, AJ, 108, 1598
\reference{Fet95}Forbes, D.A., Franx, M., \& Illingworth, G.D.\ 1995, AJ, 109, 1988
\reference{Gra01}Graham, A.W.\ 2001, AJ, 121, 820
\reference{Gra02}Graham, A.W.\ 2002a, ApJ, 568, L13
\reference{Gra02}Graham, A.W.\ 2002b, MNRAS, 334, 859
\reference{Get01}Graham, A.W., Erwin, P., Caon, N., \& Trujillo, I.\ 2001a, ApJ, 563, L11
\reference{Get01}Graham, A.W., Erwin, P., Caon, N., \& Trujillo, I.\ 2002, RevMexAA SC, in press (astro-ph/0206248)
\reference{Get96}Graham, A.W., Lauer, T.R., Colless, M.M., \& Postman, M.\ 1996, ApJ, 465, 534
\reference{GaG02}Graham, A.W., \& Guzm\'an, R.\ 2003, AJ, submitted
\reference{GTC01}Graham, A.W., Trujillo, I., \& Caon, N.\ 2001b, AJ, 122, 1707
\reference{Jam94}James, P.\ 1994, MNRAS, 269, 176
\reference{Jet94}Jaffe, W., Ford, H.C., O'Connell, R.W., van den Bosch, \& F.C., Ferrarese, L.\ 1994, AJ, 108, 1567
\reference{JaB97}Jerjen, H., \& Binggeli, B.\ 1997, in The Nature of Elliptical Galaxies; The Second Stromlo Symposium, ASP Conf.\ Ser., 116, 239
\reference{JBF00}Jerjen, H., Binggeli, B., \& Freeman, K.C.\ 2000, AJ, 119, 593
\reference{HaM}Hjorth, J., \& Madsen, J.\ 1995, 445, 55
\reference{Kee02}Keeton, C.R.\ 2002, ApJ, 582, 17
\reference{Ken87}Kent, S.M.\ 1987, AJ, 94, 306
\reference{Kin66}King, I.R.\ 1966, AJ, 71, 64
\reference{Ket03}Komossa, S., Burwitz, V., Hasinger, G., Predehl, P., Kaastra, J.S., \& Ikebe, Y.\ 2003, ApJ, 582, L15
\reference{Kor85}Kormendy, J.\ 1985, ApJ, 292, L9
\reference{Kor94}Kormendy, J., Dressler, A., Byun, Y.I., Faber, S.M., Grillmair, C., Lauer, T.R., Richstone, D., Tremaine, S.\ 1994, in Dwarf Galaxies, ESO Conference and Workshop Proceedings, ed, G.Meylan, P. Prugniel (ESO, Garching), 147
\reference{Let03}Laine, S., van der Marel, R.P., Lauer, T.R., Postman, M., O'Dea, C.P., \& Owen, F.N.\ 2003, AJ, in press (astro-ph/0211074)
\reference{Lau85}Lauer, T.R.\ 1985, ApJ, 292, 104
\reference{Let95}Lauer, T.R., et al.\ 1995, AJ, 110, 2622
\reference{Mar00}M\'arquez, I., Lima Neto, G.B., Capelato, H., Durret, F., \& Gerbal D.\ 2000, A\&A, 353, 873
\reference{Mar01}M\'arquez, I., Lima Neto, G.B., Capelato, H., Durret, F., Lanzoni, B., \& Gerbal, D.\ 2001, A\&A, 379, 767
\reference{Mak97}Makino, J.\ 1997, ApJ, 478, 58
\reference{MaE96}Makino, J., \& Ebisuzaki, T.\ 1996, ApJ, 465, 527
\reference{MaM01}Milosavljevic, M., \& Merritt, D.\ 2001, ApJ, 563, 34
\reference{MMR02}Milosavljevic, M., Merritt, D., Rest, A., \& van den Bosch, F.\ 2002, MNRAS, 331, L51
\reference{MKK01}Mu\~noz, J.A., Kochanek, C.S., \& Keeton, C.R.\ 2001, ApJ, 558, 657 
\reference{Pet96}Phillips, A.C., Illingworth, G.D., MacKenty, J.W., \& Franx, M.\ 1996, AJ, 111, 1566
\reference{QBS00}Quillen, A.C., Bower, G.A., \& Stritzinger, M.\ 2000, ApJS, 128. 85
\reference{RHP02}Ravindranath, S., Ho, L.C., \& Filippenko, A.V.\ 2002, ApJ, 566, 801
\reference{Rav01}Ravindranath, S., Ho, L.C., Peng, C.Y., Filippenko, A.V., \& Sargent, W.L.W.\ 2001.\ AJ, 122, 653
\reference{Ret01}Rest, A., et al.\ 2001, AJ, 121, 2431 
\reference{Sch79}Schweizer, F., 1979, ApJ, 233, 23
\reference{Ser68}S\'ersic, J.-L.\ 1968, Atlas de Galaxias Australes (Cordoba: Observatorio Astronomico) 
\reference{Set02}Seigar, M.S., Carollo, C.M., Stiavelli, M., de Zeeuw, P.T., \& Dejonghe, H.\ 2002, AJ, 123, 184
\reference{SaJ98}Seigar, M.S., \& James, P.A.\ 1998, MNRAS, 299, 672
\reference{Set01}Stiavelli, M., Miller, B.W., Ferguson, H.C., Mack, J., Whitmore, B.C., \& Lotz, J.M.\ 2001, AJ, 121, 1385
\reference{Ton84}Tonry, J.L.\ 1984, ApJ, 283, L27
\reference{Tet03}Trujillo, I., Erwin, P., Asensio-Ramos, A., Graham, A.W.\ 2003, (Paper II) in prep
\reference{TGC01b}Trujillo, I., Graham, A.W., \& Caon, N.\ 2001, MNRAS, 326, 869
\reference{vdM99}van der Marel, R.P.\ 1999, AJ, 117, 744
\reference{VHM03}Volonteri, M., Haardt, F., \& Madau, P.\ 2003, ApJ, 582, 559
\reference{YaK94}Young C.K., \& Currie M.J.\ 1994, MNRAS, 268, L11 
\reference{Hon96}Zhao, H.S.\ 1996, MNRAS, 278, 488
\reference{Hon96}Zhao, H.S.\ 1997, MNRAS, 287, 525 
\reference{ZHR02}Zhao, H.S.\ Haehnelt, M.G., \& Rees, M.J.\ 2002, New Astronomy, 7, 385
\end{references}
\end{document}